\documentclass[twocolumn,superscriptaddress,prc]{revtex4}
\usepackage{graphicx}
\usepackage{diagbox}
\usepackage{bm}
\usepackage[colorlinks=true,linkcolor=blue, citecolor=blue]{hyperref}

\begin{document}
\title{A transport model study of multiparticle cumulants in $p+p$ collisions at 13 TeV}
\author{Xin-Li Zhao}
\affiliation{Key Laboratory of Nuclear Physics and Ion-beam
Application (MOE), Institute of Modern Physics, Fudan University,
Shanghai 200433, China}
\affiliation{Department of Physics, East Carolina University, Greenville, North Carolina 27858, USA}
\author{Zi-Wei Lin}
\email[]{linz@ecu.edu}
\affiliation{Department of Physics, East Carolina University, Greenville, North Carolina 27858, USA}
\author{Liang Zheng}
\affiliation{School of Mathematics and Physics, China University of Geosciences (Wuhan), Wuhan 430074, China}
\author{Guo-Liang Ma}
\affiliation{Key Laboratory of Nuclear Physics and Ion-beam
Application (MOE), Institute of Modern Physics, Fudan University,
Shanghai 200433, China}

\begin{abstract}
Flow-like signals including the ridge structure observed in small collision systems that are similar to those in large collision systems have led to questions about the onset of collectivity in nuclear collisions. In this study, we use the string melting version of a multi-phase transport (AMPT) model with or without the sub-nucleon geometry for  the proton to study multiparticle cumulants in $p+p$ collisions at 13 TeV. 
Both versions of the model produce negative $c_{2}\{4 \}$ values at high multiplicities. 
In addition, the dependence of $c_{2}\{4 \}$ on the parton cross section is non-monotonous, where only a range of parton cross section values leads to   negative  $c_{2} \{4  \}$. 
Furthermore, the AMPT model with sub-nucleon geometry better describes the multiplicity dependence of $c_{2} \{4  \}$, demonstrating the importance of incorporating the sub-nucleon geometry in studies of small collision systems. 
\end{abstract}

\maketitle

\section{Introduction}
\label{introduction}

The quark-gluon plasma (QGP) with deconfined parton degrees of freedom is created at relativistic heavy ion collisions such as those at the Relativistic Heavy Ion Collider (RHIC) and the Large Hadron Collider (LHC). Many final state observables, such as hadron spectra, collective flows and fluctuations, can be sensitive to the formation of the QGP. In particular, the long-range ``ridge'' structures in two-particle azimuthal correlations observed in large system A+A collisions~\cite{PHENIX:2008osq,STAR:2009ngv,ALICE:2011ab,CMS:2013wjq}, small system A+A collisions~\cite{PHENIX:2013ktj,PHENIX:2017xrm,PHENIX:2018lia} and the even smaller $p$+A or $p+p$ systems ~\cite{ATLAS:2015hzw,CMS:2015yux,ATLAS:2016yzd} are strikingly similar. 

The ridge structure in large system A+A collisions is generally thought to be produced by the hydrodynamic expansion \cite{Ollitrault:1992bk,Heinz:2013th}
or multiple parton collisions \cite{Lin:2001zk,Xu:2007jv,Alver:2010gr}
of the hot and dense matter. 
For small systems, however, hydrodynamics-based models are expected to be different from transport models because there are not many rescatterings per parton or per hadron. Indeed, a non-equilibrium parton escape mechanism is found to dominate the development of anisotropic flows for collisions at energies that are not high enough and for small systems~\cite{He:2015hfa,Lin:2015ucn}. 
A key question about the ridge is whether it is due to collective flow (i.e., many particles that correlate with a common event plane) or nonflow such as resonance decays and momentum conservation. 
The multiparticle cumulant method has been applied to $p$+Pb and $p+p$ collisions~\cite{Zhao:2017rgg,ATLAS:2017hap,ATLAS:2017rtr,Bzdak:2017zok,Nie:2018xog} to suppress the nonflow effects and better extract the flow signals. 
For example, a negative four-particle cumulant $c_{2} \{4  \}$ is expected when the correlation comes from the collective flow~\cite{Borghini:2001vi,Jia:2014pza,Jia:2017hbm}. 

A hydrodynamics-based hybrid model has been used to investigate the $c_{2} \{4  \}$ in $p+p$ collisions, and $c_{2} \{4  \}>0$ is found from different analysis methods including the standard cumulants, two-subevent and three-subevent cumulants~\cite{Zhao:2017rgg}. 
In this work, we apply a multi-phase transport (AMPT) model~\cite{Lin:2004en,Zhang:2019utb,Zhang:2021vvp} to study the  multiparticle cumulants including  $c_{2} \{4  \}$ using  both the standard cumulants and three-subevent cumulants. 

The paper is organized as follows. First, we introduce the AMPT model used in this study in  Sec.~\ref{ampt}. Then the multiparticle cumulant methods including the standard cumulants and the subevent cumulants are described in Sec.~\ref{cumulant}. In Sec.~\ref{results}, we present the results and discussions. Finally, conclusions are given in Sec.~\ref{summary}.

\section{A multi-phase transport model}
\label{ampt}

The string melting version of the AMPT model~\cite{Lin:2001zk,Lin:2004en} contains four main parts to describe nuclear collisions: a fluctuating initial condition from the HIJING model, elastic parton scatterings from the Zhang's parton cascade, hadronization from a quark coalescence model, and hadronic scatterings  based on the ART model. 
The model is able to reasonably describe the collective flow from small to and large systems at RHIC and LHC energies~\cite{Bzdak:2014dia,Ma:2016fve,Li:2018leh,Wei:2018xpm,Nie:2018xog}.

In this work, we use a recently developed version of the AMPT model, which uses a new quark coalescence model~\cite{He:2017tla}, a modern set of parton distributions functions in the proton and an impact parameter-dependent nuclear shadowing~\cite{Zhang:2019utb}, 
and improved heavy quark productions~\cite{Zheng:2019alz,Lin:2021mdn}. 
This model has been shown to work well in describing the particle yields and $p_{\rm T}$ spectra in $p+p$ and A+A collisions at high energies. 
Note that for this work on $p+p$ collisions we take the same values for the Lund string fragmentation parameters: $a_L=0.8$ and $b_L=0.4$ GeV$^{-2}$~\cite{Zhang:2019utb, Zheng:2019alz}. 

In addition to the above normal AMPT model, where the proton is treated as a point particle, 
we also use a modified AMPT model that includes the sub-nucleon geometry of the proton following a recent study~\cite{Zheng:2021jrr}. 
The sub-nucleon geometry has been shown to lead to different spatial fluctuations and affect the collective flow of the small systems~\cite{Schenke:2014zha,Mantysaari:2016ykx,Welsh:2016siu,Mantysaari:2017cni}. 
The matter distribution of the proton, based on the proton charge form factor, is given by
$\rho(r) \propto e^{-r/R}$ with $R=0.2$ fm~\cite{Zheng:2021jrr}.  
In the constituent quark picture, a proton is assumed to consist of three constituent quarks, which coordinates are sampled according to the matter distribution. 
Then the proton-proton collision can be extended to the participant quark geometries within  
the Glauber model framework~\cite{Zheng:2021jrr}. 
In this study, we name the AMPT version with sub-nucleon geometry of the proton as ``3-quark AMPT'' while the version without the sub-nucleon geometry is named ``normal AMPT''. For both versions, the parton cross section is given by $\sigma=4.5\pi\alpha_{s}^{2}/\mu^{2}$, where the parameter $\mu$ represents the Debye screening mass~\cite{Lin:2004en}.

\section{Multiparticle Cumulants}
\label{cumulant}

An advantage of the multiparticle cumulant method is that it suppresses nonflow effects  such as those from jets and dijets. Recently the method has been widely applied to A+A collisions and small system collisions. 
The multiparticle cumulant method using moments of $Q$ vectors is called the standard cumulants, direct cumulants, or $Q$ cumulants~\cite{Bilandzic:2010jr}, where $Q_{n} \equiv  \sum_i e^{in\phi _{i}}$ are the flow vectors. 
The extended method of subevent cumulants can further suppress the nonflow effects~\cite{Jia:2017hbm}.

We calculate the two-particle cumulants using the standard method and four-particle cumulants using both the standard and three-subevent methods. 
In the standard cumulant method, two-particle and four-particle cumulants are given by 
\begin{eqnarray}
\langle  \langle  \{2  \}_{n} \rangle  \rangle&=&  \langle  \langle e^{in(\phi _{1}-\phi _{2})}  \rangle  \rangle, \nonumber \\
\langle  \langle  \{4  \}_{n} \rangle  \rangle&=&  \langle  \langle e^{in(\phi _{1}+\phi _{2}-\phi _{3}-\phi _{4})}  \rangle  \rangle,
 \label{dsigmadt}
\end{eqnarray}
in which the double brackets mean the weighted averaging over all particles in an event and then over all events. We then have
\begin{eqnarray}
c_{n} \{2  \}&=& \langle  \langle   \{2  \}_{n}  \rangle  \rangle, \nonumber \\ 
c_{n} \{4  \}&=& \langle  \langle   \{4  \}_{n}  \rangle  \rangle-2 \langle  \langle   \{2  \}_{n}  \rangle  \rangle ^{2} .
\label{eqc24}
\end{eqnarray}

In the subevent method, the overall events are organized into multiple subevents according to pseudorapidity $\eta$, where each subevent covers a non-overlapping $\eta$ interval. 
In particular, in the three-subevent method, the overall events are divided into three subevents: $a$ within $-\eta_{\mathrm{max}}<\eta<-\eta_{\mathrm{max}}/3$, $b$ within $-\eta_{\mathrm{max}}/3 <\eta<\eta_{\mathrm{max}}/3$, and $c$ within $\eta_{\mathrm{max}}/3<\eta<\eta_{\mathrm{max}}$. 
Then the corresponding four-particle cumulants are defined as
\begin{eqnarray}
\langle   \langle \{4  \}_{n}  \rangle \rangle_{\rm three-sub}
=\langle  \langle e^{in(\phi _{1}^{a}+\phi _{2}^{a}-\phi _{3}^{b}-\phi _{4}^{c})}  \rangle \rangle.
\end{eqnarray}

In the comparisons with the ATLAS data, we use $\eta _{\mathrm{max}}=2.5$ to match the ATLAS detector coverage. Specifically, the multiparticle cumulants are calculated in three steps~\cite{ATLAS:2017hap,ATLAS:2017rtr}. First, multiparticle correlations $\langle  \{2k  \}_{n}  \rangle$ are calculated for reference (charged) particles with $0.3<p_{\rm T}<3$ GeV$/c$ for each event. 
Second, these $ \langle  \{2k  \}_{n}  \rangle$ correlations are averaged over all events with the same $N_{\rm ch}^{\rm sel}$, which represents the number of charged particles within a given $p_{\rm T}$ range ($0.3<p_{\rm T}<3$ GeV$/c$ being the default range, plus $p_{\rm T}>0.2$ GeV$/c$, $p_{\rm T}>0.4$ GeV$/c$, or $p_{\rm T}>0.6$ GeV$/c$). This leads to the corresponding $c_{n} \{2k  \}$ values that are averaged over events with the same $N_{\rm ch}^{\rm sel}$. Third, the $c_{n} \{2 \}$ and $c_{n} \{4 \}$ values for a given range of $N_{\rm ch}^{\rm sel}$ are mapped to 
the final $c_{n} \{2 \}$ value and $c_{n} \{4 \}$ value at the corresponding $N_{\rm ch}$ ($p_{\rm T}>0.4$ GeV$/c$) value. 
Note that all results of $c_{n} \{2k  \}$ shown in this study are for reference particles within $0.3<p_{\rm T}<3$ GeV$/c$, where the standard cumulant method and the $p_{\rm T}$ range  $0.3<p_{\rm T}<3$ GeV$/c$ for $N_{\rm ch}^{\rm sel}$ are used unless stated otherwise. 
Typically we simulate about 150 million $p+p$ events for each case.

\section{$c_{2} \{2 \}$ \& $c_{2} \{4 \}$ results}
\label{results}

\begin{figure*}[htb]
\includegraphics[width=1.\textwidth]{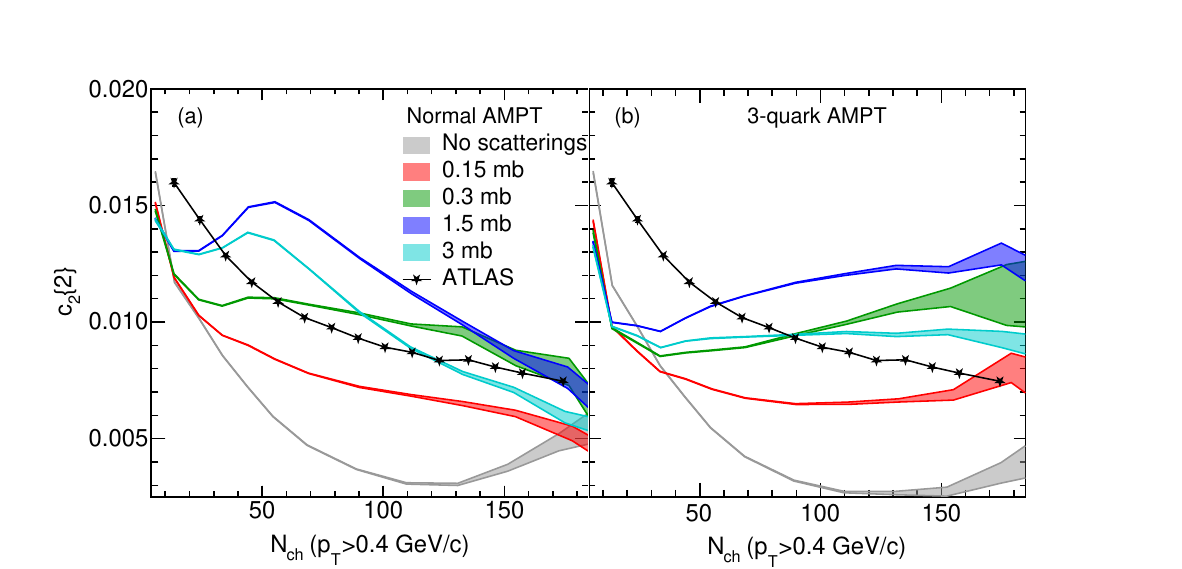}
\caption{$c_{2} \{2  \}$ for reference particles within $0.3<p_{\rm T}<3$ GeV$/c$ and $|\eta|<2.5$ from (a) the normal AMPT model and (b) the 3-quark AMPT model in comparison with the ATLAS data; the event averaging uses $N_{\rm ch}^{\rm sel}$ within $0.3<p_{\rm T}<3$ GeV$/c$.}
\label{fig:c22}
\end{figure*}

Figure~\ref{fig:c22} shows the comparison of the $c_{2} \{2 \}$ results from the AMPT model 
with different parton cross sections in comparison with the ATLAS data~\cite{ATLAS:2017hap}. 
Note that the standard cumulant method is used here, both reference particles and particles used for $N_{\rm ch}^{\rm sel}$ are within $0.3<p_{\rm T}<3$ GeV$/c$, 
and the AMPT results without (left panel) and with (right panel) the sub-nucleon structure for the proton are both shown. 
We see that $c_{2} \{2  \}$ is very sensitive to the value of the parton scattering cross section $\sigma$. 
When $\sigma$ is set to zero and the hadron cascade is also turned off, 
the AMPT model has no secondary scatterings and gives the black curves in Fig.~\ref{fig:c22}, which in both panels are lower than the $c_{2} \{2  \}$ data. 
On the other hand, at certain $\sigma$ values $c_{2} \{2  \}$ from the AMPT model can be larger than the data at the same multiplicity. 
The $\sigma$-dependence of the results also shows a surprising behavior.
For example, within the multiplicity range $N_{\rm ch}(p_{\rm T}>0.4{\rm~GeV}/c) \in (40,150)$, the $c_{2}  \{2  \}$ value first increases with the parton cross section and then decreases once $\sigma \ge 1.5$ mb. 

We also see that the sub-nucleon structure has a significant effect on the multiplicity dependence 
and $\sigma$-dependence of $c_{2} \{2  \}$. In Fig.~\ref{fig:c22}(a), the $c_{2} \{2  \}$ values at 0.3  mb or 3 mb from the AMPT model without the sub-nucleon structure are relatively close to the ATLAS data; while in Fig.~\ref{fig:c22}(b) the $c_{2} \{2  \}$ values at 3 mb from the AMPT model  with the sub-nucleon structure are relatively close to data. However, none of the results at a constant $\sigma$ can well reproduce the ATLAS $c_{2} \{2  \}$ data. 

\begin{figure*}[htb]
\includegraphics[width=1.\textwidth]{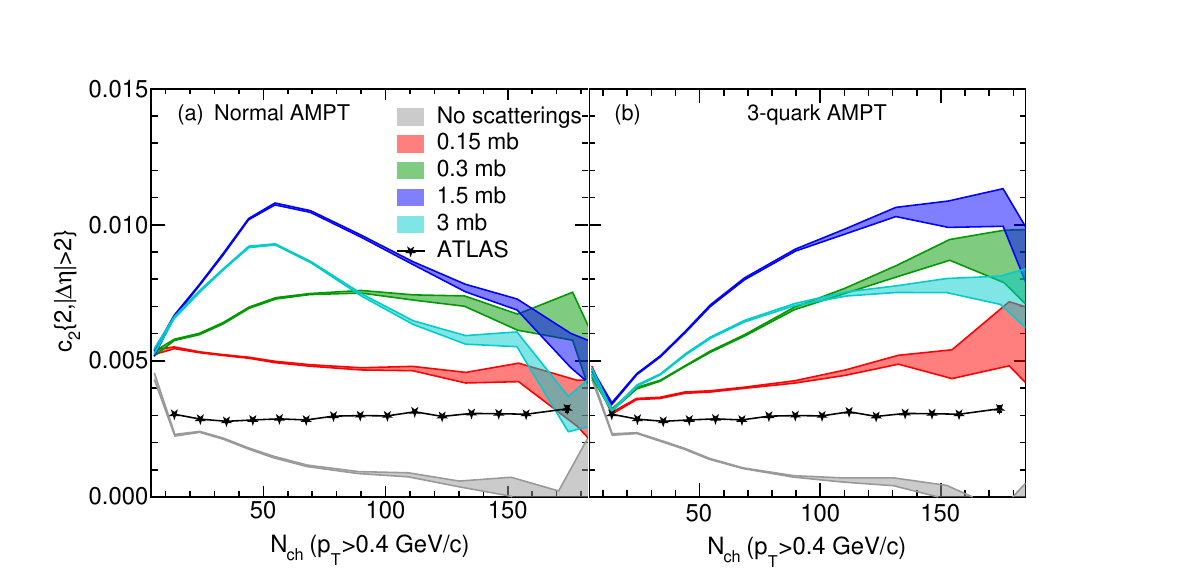}
\caption{Same as Fig.~\ref{fig:c22}, but for $c_{2} \{2 ,|\Delta\eta|>2 \}$.}
\label{fig:c22d}
\end{figure*}

To further suppress the nonflow effect, a separation in pseudorapidity of $|\Delta\eta|>2$ for the two hadrons forming a pair is applied to the $c_{2} \{2  \}$ calculation, and the corresponding model results are presented in Fig.~\ref{fig:c22d} in comparison with 
the experimental data. 
We see that the $c_{2} \{2 ,|\Delta\eta|>2 \}$ values are much smaller than the corresponding 
$c_{2} \{2\}$ values at low $N_{\rm ch}$, which is also the case for the experimental data. 
This means that the nonflow effect is especially significant at low multiplicities. 
On the other hand, the decrease of $c_{2} \{2\}$ due to the pseudorapidity gap 
from the AMPT model can be quite different from that in the data. For example, 
the AMPT results at $\sigma=0.15$ mb from both the normal AMPT and 3-quark AMPT are mostly below the ATLAS data in Fig.~\ref{fig:c22} but above the data in Fig.~\ref{fig:c22d}.
This shows that the AMPT model does not have the correct nonflow~\cite{PHENIX:2017xrm}.
Again, none of the model results at a constant $\sigma$ can well reproduce the ATLAS $c_{2} \{2 ,|\Delta\eta|>2 \}$ data. 
We have also checked the differential flow $v_{2} (p_{\rm T} )$, which results are consistent with the $c_{2} \{2  \}$ results including similar deviations from the experimental data.

\begin{figure*}[htb]
\includegraphics[width=1.\textwidth]{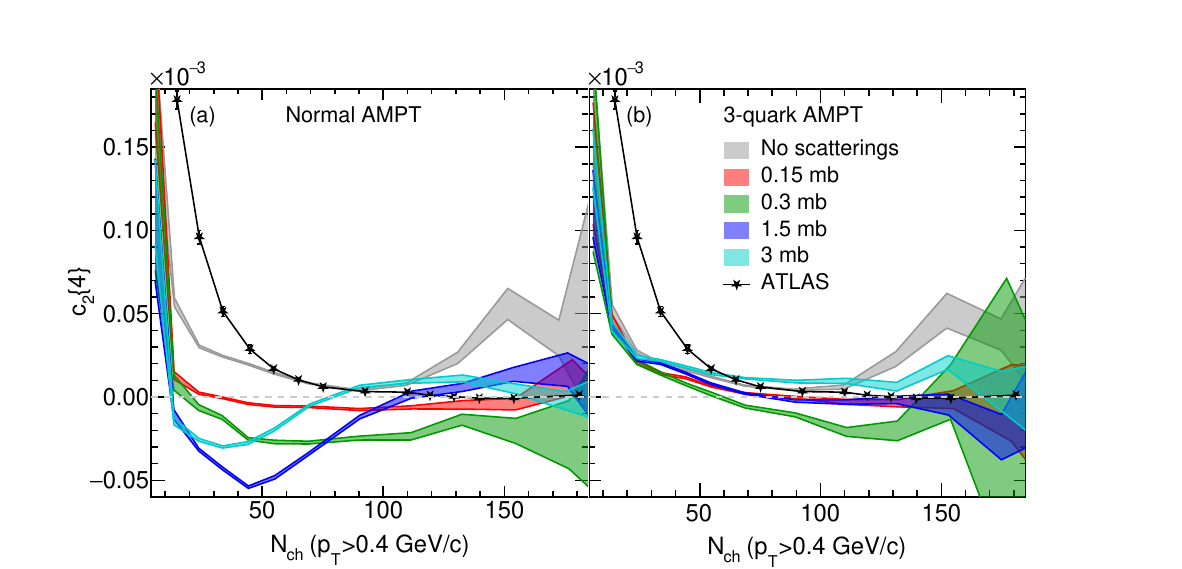}
\caption{$c_{2} \{4 \}$ for reference particles within $0.3<p_{\rm T}<3$ GeV$/c$ and $|\eta|<2.5$ from (a) the normal AMPT model and (b) the 3-quark AMPT model in comparison with the ATLAS data.} 
\label{fig:c24}
\end{figure*}

In Fig.~\ref{fig:c24}, we show the $c_{2} \{4  \}$ results 
from the standard cumulant method at different parton cross sections. 
First, we see that  the $c_{2} \{4  \}$ values from the AMPT model without any parton or hadron scatterings (black curves) are all positive for both the normal and  3-quark AMPT models.  Also, all the $c_{2} \{4  \}$ values at very low multiplicities (i.e., $N_{\rm ch}$) are  positive, reflecting the contribution from nonflow effects such as the global momentum conservation
~\cite{Bzdak:2017zok}. 
When the parton cross section is non-zero,  the $c_{2} \{4  \}$ values at high multiplicities are often negative. In Fig.~\ref{fig:c24}(a), the normal AMPT model $\sigma=0.15$ or 0.3 mb  both produces mostly negative $c_{2} \{4  \}$ values when $N_{\rm ch}>32$, 
quite different from the ATLAS data~\cite{ATLAS:2017rtr} that show the sign change at $N_{\rm ch} \sim 130$. 
At $\sigma=1.5$ or 3 mb, $c_{2} \{4  \}$ from the normal AMPT model 
has large negative values at relatively low multiplicities before becoming positive; 
this multiplicity dependence is very different from the ATLAS data. 

On the other hand, the $c_{2} \{4  \}$ results from the 3-quark AMPT model in Fig.~\ref{fig:c24}(b) show a similar multiplicity dependence as the ATLAS data. 
We see that the $c_{2} \{4  \}$ values at $\sigma=0.15$ mb or $\sigma=1.5$ mb are  close to each other and both become negative at $N_{\rm ch} \sim 80$. 
It is also obvious that $c_{2} \{4  \}$ has a non-monotonous dependence on the parton cross section $\sigma$. For example, the $c_{2} \{4  \}$ value at $N_{\rm ch} \sim 100$ first decreases with $\sigma$ and becomes negative, and then it increases with $\sigma$ and becomes positive again. This is the case for both the normal and  3-quark AMPT models.

\begin{figure*}[htb]
\includegraphics[width=1.\textwidth]{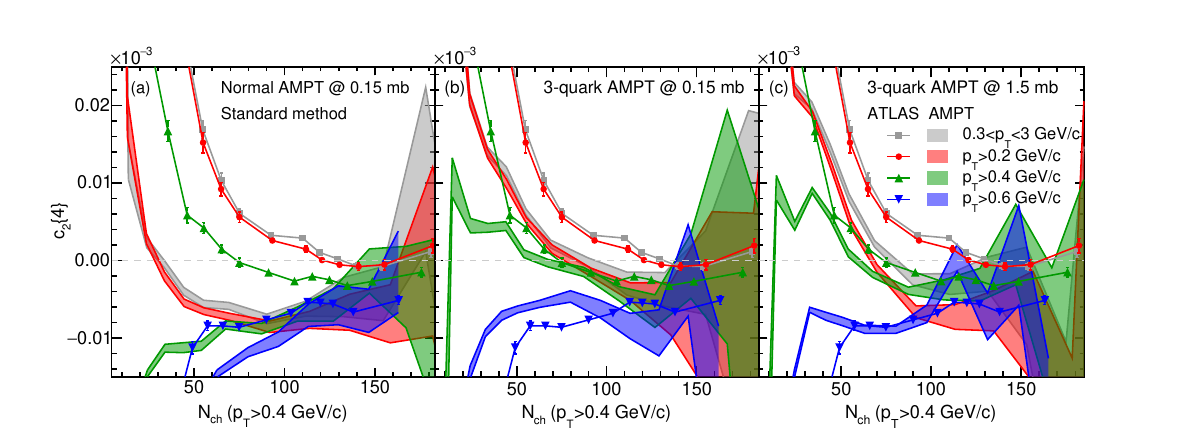}
\caption{$c_{2} \{4  \}$ results from (a) the normal AMPT model at 0.15 mb, (b) the 3-quark AMPT model at 0.15 mb, and (c) the 3-quark AMPT at 1.5 mb when different $p_{\rm T}$ ranges are used for $N_{\rm ch}^{\rm sel}$ in the event averaging.}
\label{fig:npt4}
\end{figure*}

In Fig.~\ref{fig:npt4}, we choose different $p_{\rm T}$ ranges for the calculation of $N_{\rm ch}^{\rm sel}$ in the event averaging procedure and compare the results with the ATLAS data~\cite{ATLAS:2017rtr}.
Results from the normal AMPT model for $\sigma=0.15$ mb are shown in Fig.~\ref{fig:npt4}(a), while results from the 3-quark AMPT model for $\sigma=0.15$ and 1.5 mb are shown in Fig.~\ref{fig:npt4}(b) and (c), respectively. 
In all three cases, we see that the model results and the ATLAS data show the same qualitative behavior, in that $c_{2} \{4  \}$ at low $N_{\rm ch}$ decreases significantly with the increase of  the minimum $p_{\rm T}$ used for $N_{\rm ch}^{\rm sel}$ while $c_{2} \{4  \}$ at intermediate $N_{\rm ch}$ changes much less. 
Also, the $c_{2} \{4  \}$ values for the $p_{\rm T}$ range $p_{\rm T}>0.6$ GeV/$c$ are 
mostly negative in both the model results and the experimental data. 
Looking more closely, we see that the results from the normal AMPT model can be quite different from the 3-quark AMPT results. 
For example, the $c_{2} \{4  \}$ curve for the $p_{\rm T}$ range $p_{\rm T}>0.4$ GeV/$c$ 
in Fig.~\ref{fig:npt4}(a) from the normal AMPT model are almost all negative, while the $c_{2} \{4  \}$ values from the-3 quark AMPT model in Fig.~\ref{fig:npt4}(b) and (c) change from  positive to negative at $N_{\rm ch} \sim 60$ similar to the data.

\begin{figure*}[htb]
\includegraphics[width=1.\textwidth]{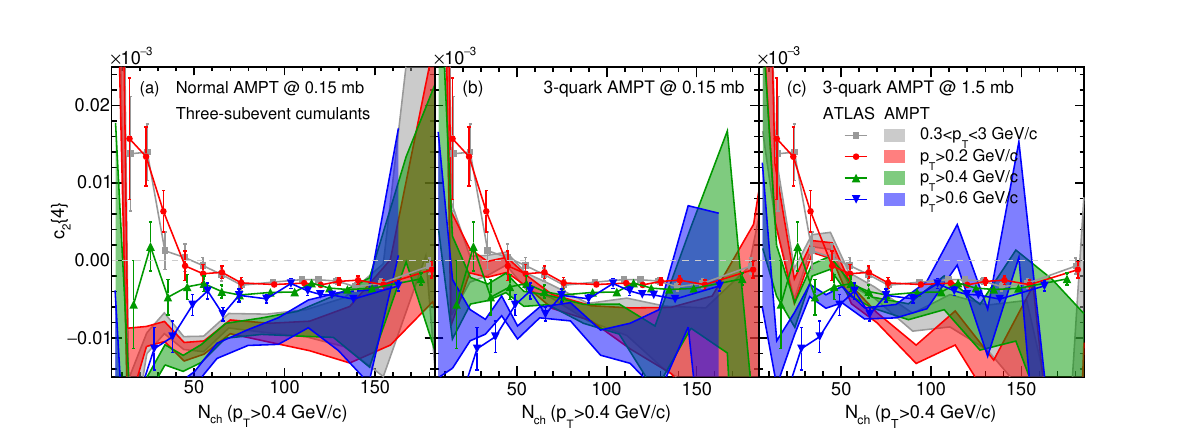}
\caption{Same as Fig.~\ref{fig:npt4}, but using the three-subevent method.}
\label{fig:3sub}
\end{figure*}

\begin{figure*}[htb]
  \begin{minipage}[t]{0.5\linewidth}
    \centering
    \includegraphics[width=1.0\textwidth]{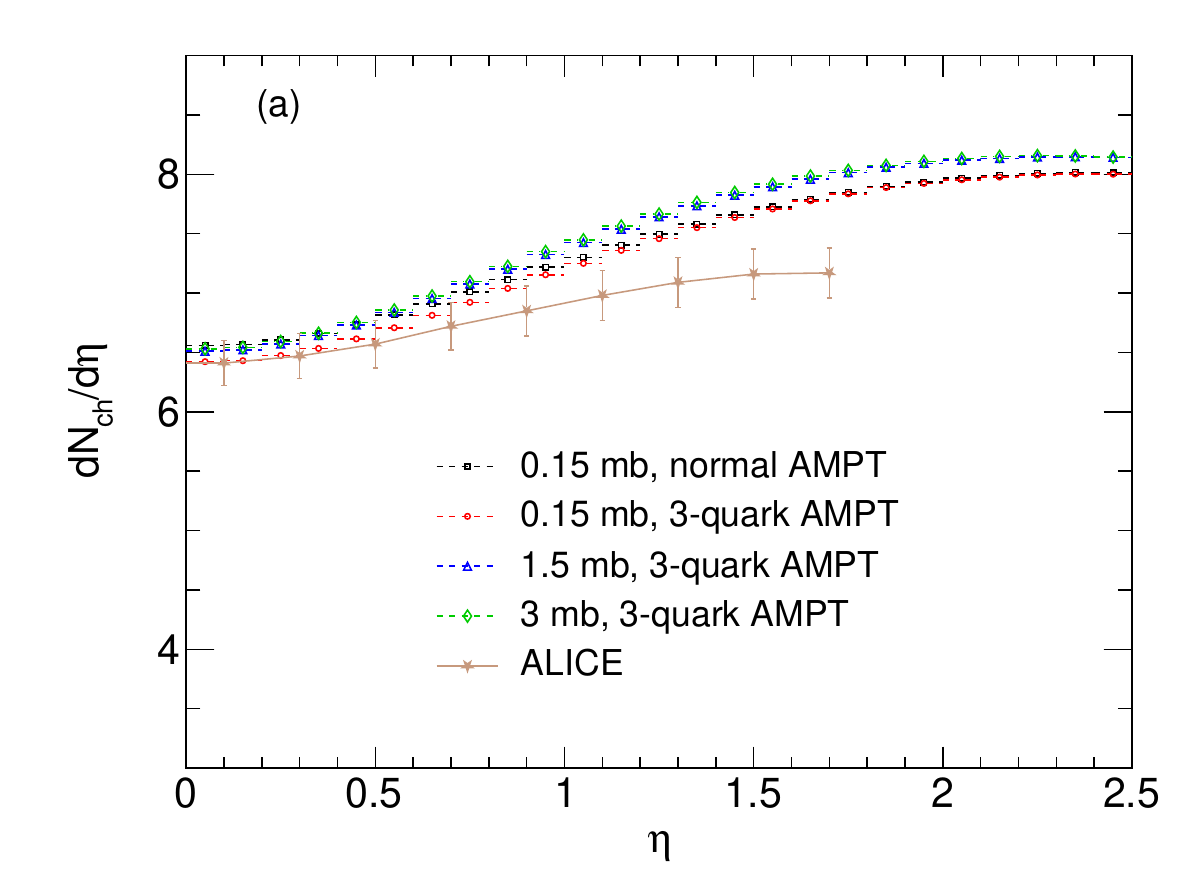}
  \end{minipage}%
     \begin{minipage}[t]{0.5\linewidth}
    \centering
    \includegraphics[width=1.0\textwidth]{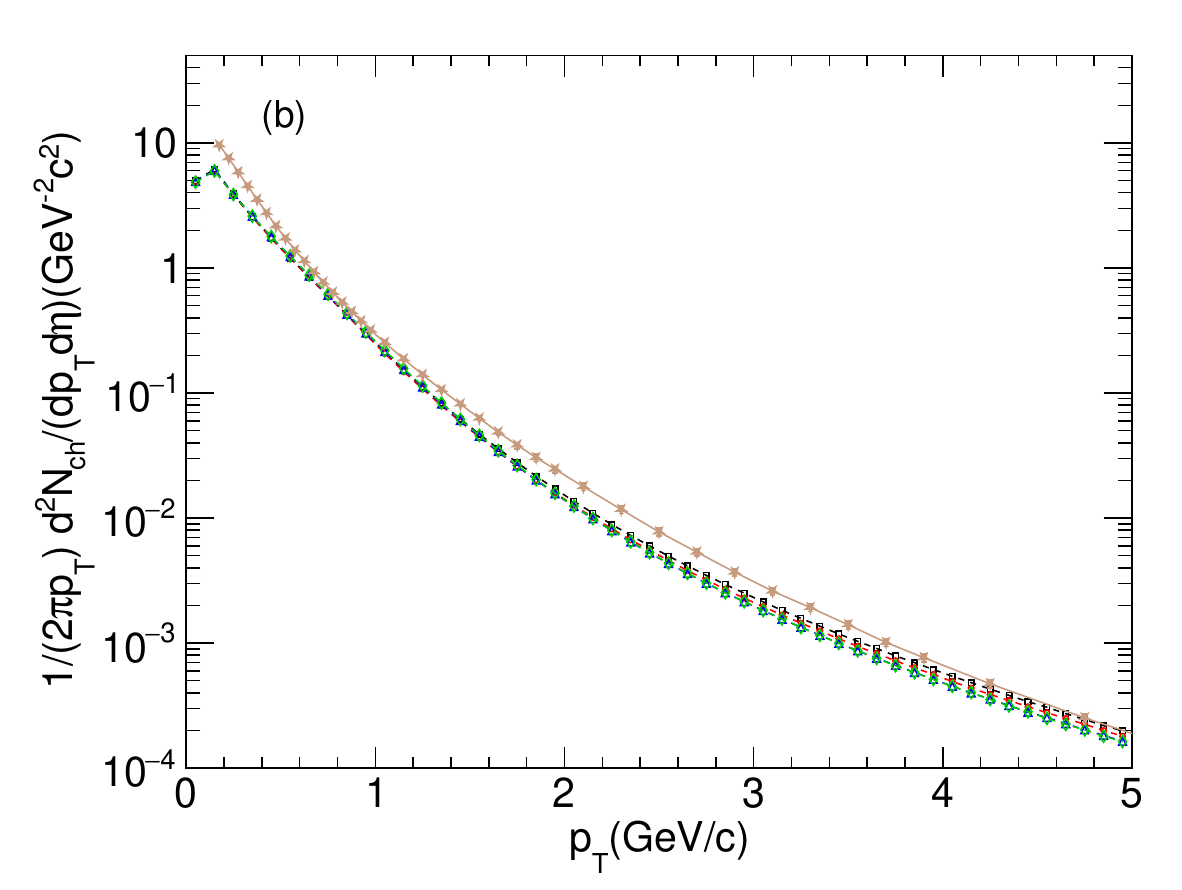}
  \end{minipage}
\caption{(a) The pseudorapidity distributions and (b) the $p_{\rm T}$ spectra of charged particles 
from the normal AMPT and 3-quark AMPT models in comparison with the ALICE data~\cite{ALICE:2015qqj}.}
\label{fig:pt}
\end{figure*}

\begin{figure*}[htb]
\includegraphics[width=0.5\textwidth]{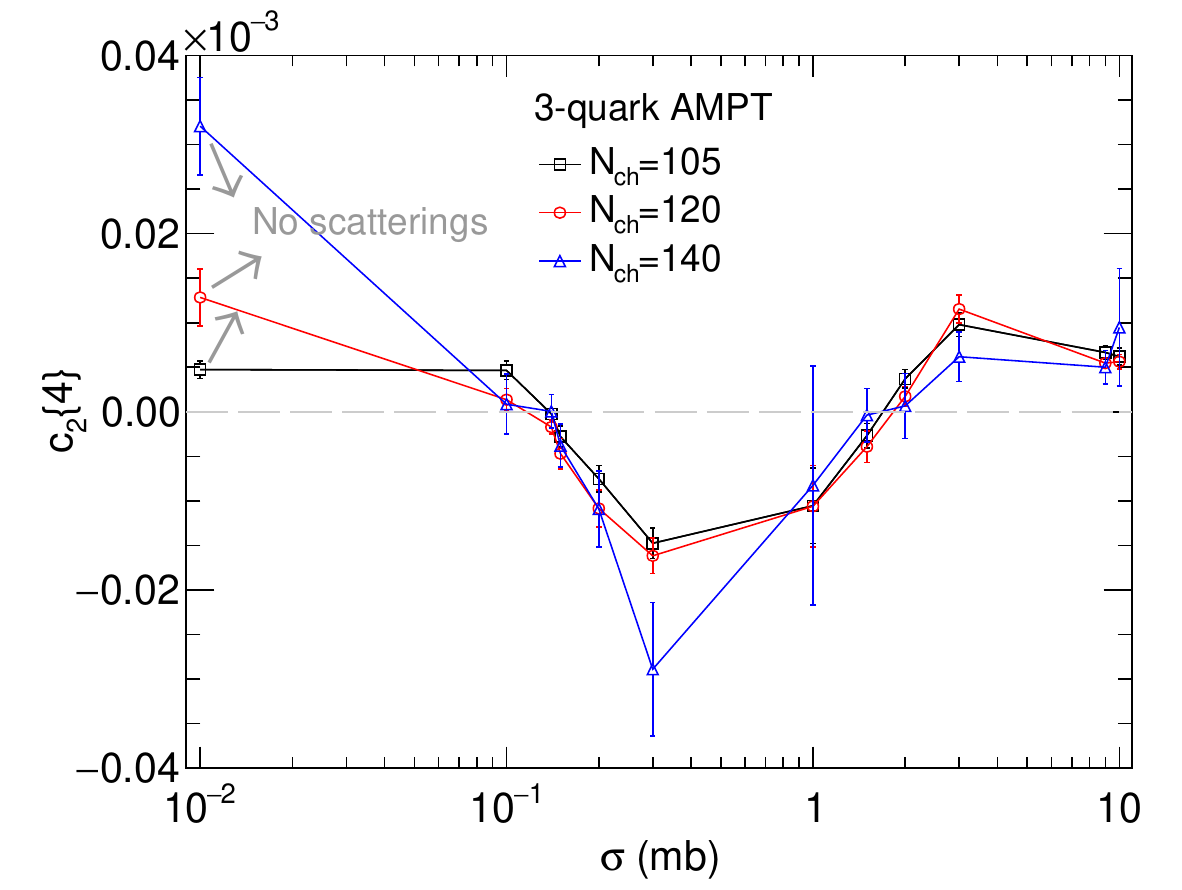}
\caption{$c_{2} \{4  \}$ from the 3-quark  AMPT model at different multiplicities versus the parton cross section, where the standard cumulant method is used.}
\label{fig:sigma}
\end{figure*}

\begin{figure*}[htb]
    \begin{minipage}[t]{0.5\linewidth}
    \centering
    \includegraphics[width=1.0\textwidth]{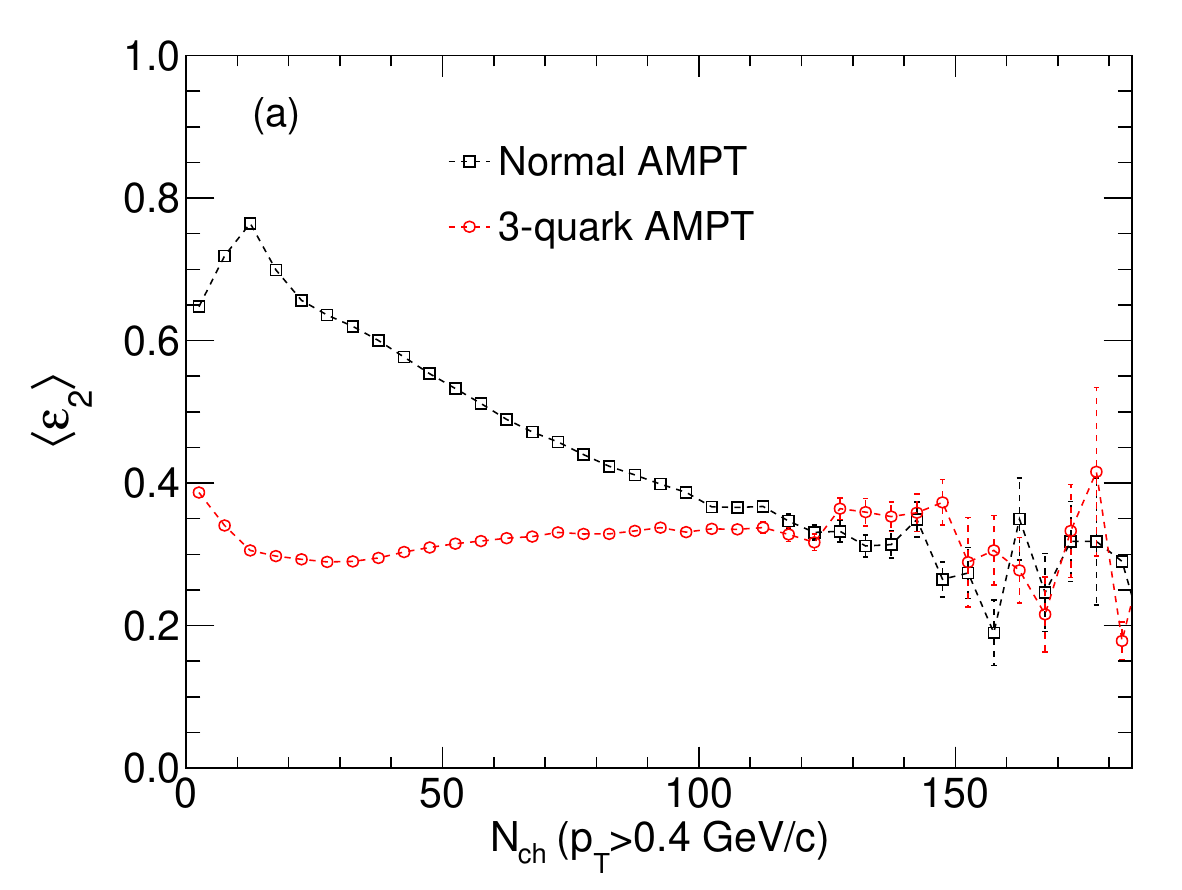}
  \end{minipage}%
   \begin{minipage}[t]{0.5\linewidth}
    \centering
    \includegraphics[width=1.0\textwidth]{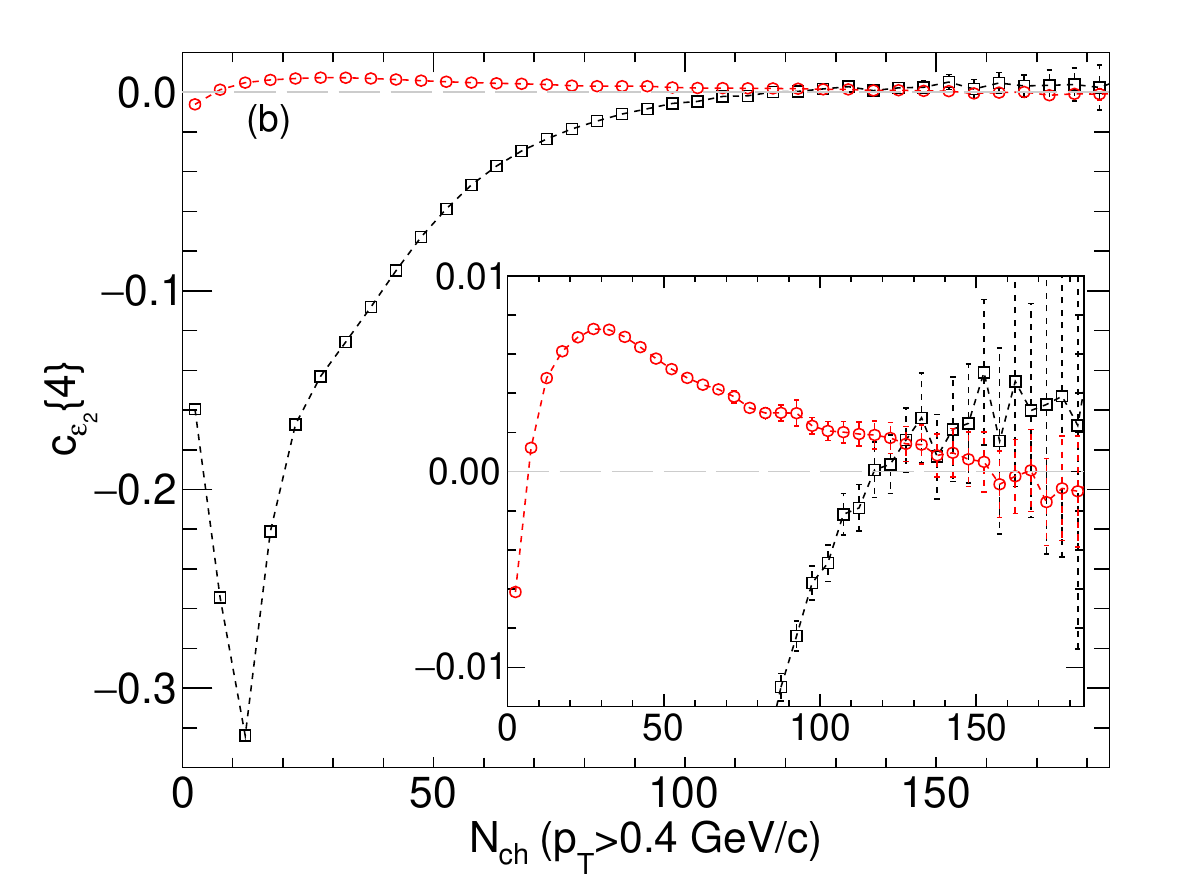}
  \end{minipage}
\caption{(a) Event-averaged spatial eccentricity of initial partons and  (b) $c_{\varepsilon_2} \{4  \}$ versus $N_{\rm  ch}$ from the normal AMPT and 3-quark AMPT models.}
\label{fig:e2}
\end{figure*}

To suppress the nonflow effects, we also apply the three-subevent cumulant method, 
and the corresponding $c_{2} \{4  \}$ results for different $p_{\rm T}$ ranges for the calculation of $N_{\rm ch}^{\rm sel}$ are shown in Fig.~\ref{fig:3sub} in comparison with the ATLAS  data~\cite{ATLAS:2017rtr}. 
Compared with the results from the standard cumulant method shown in Fig.~\ref{fig:npt4},   the decrease of $c_{2} \{4  \}$ with the increasing $p_{\rm T}$ cut used for $N_{\rm ch}^{\rm sel}$ is generally less. In addition, more $c_{2} \{4  \}$ values from the  three-subevent cumulant method are negative. These features can be observed from both the AMPT model results and  the ATLAS data. 
Although the $c_{2} \{4  \}$ magnitudes from the AMPT model 
are often quite different from the experimental data, 
our results show that $c_{2} \{4  \}$ including its sign change location in $N_{\rm ch}$ depends sensitively on the sub-nucleon geometry for the proton, and generally the 
3-quark AMPT model that includes the sub-nucleon geometry perform better than the normal AMPT model without the sub-nucleon geometry.

\section{Discussions}
\label{sec:dis}

The AMPT model that we are using here is able to reasonably describe the charged particle production  ~\cite{Zhang:2019utb,Zhang:2021vvp}. In Fig.~\ref{fig:pt}, we compare with the ALICE data for inelastic events (INEL$>$0), i.e., events having at least one charged particle within  $|\eta|<1$~\cite{ALICE:2015qqj}. Figure~\ref{fig:pt}(a) shows the pseudorapidity distributions of charged particles from the normal AMPT model at $\sigma=0.15$ mb and the 3-quark AMPT model at 0.15 mb, 1.5 mb and 3 mb, where all the model results are close to each other and similar to the data. 
Figure~\ref{fig:pt}(b) shows the $p_{\rm T}$ spectra of charged particles within $|\eta|<0.8$; 
the model results are close to each other and follow the data qualitatively, but there are some 
quantitative differences from the experimental data.

In Fig.~\ref{fig:sigma}, we further check the non-monotonous dependence of $c_{2} \{4   \}$ on the parton cross section $\sigma$, where we show the results from the 3-quark AMPT model with the standard cumulant method versus $\sigma$ at three multiplicities. 
Note that the points plotted at $\sigma=0.01$ mb actually represent
the AMPT results where both parton and hadron rescatterings are turned off. 
Because the sign change in the ATLAS data appears at $N_{\rm ch}\sim 130$~\cite{ATLAS:2017rtr}, we show the $c_{2} \{4  \}$ results around that $N_{\rm ch}$ value. 
We see that for all three multiplicities the $c_{2} \{4  \}$ values cross zero 
at $\sigma \simeq 0.15$ mb and at $\sigma \simeq 1.5$ mb, where $c_{2} \{4  \}$ is negative when the parton cross section is between these two values. 
It is interesting to see that $c_{2} \{4  \}$ is so sensitive to the parton cross section as it changes sign at a very small $\sigma$ value. 

We have seen from Figs.~\ref{fig:c22} and \ref{fig:c22d} that 
the multiplicity dependence of $c_{2} \{2 \}$ in the normal AMPT model is 
quite different from that in the 3-quark AMPT model. 
In addition, Fig. ~\ref{fig:c24} shows that 
the $c_{2} \{4  \}$ shapes from the two models are often quite different. 
For example, $c_{2} \{4  \}$ from the normal AMPT model has large negative values around   $N_{\rm ch} \sim 40$, which is not the case for the 3-quark AMPT model. 
We can partially understand these differences from the spatial eccentricity $\varepsilon_2$, 
which is calculated using the spatial distribution of all initial partons~\cite{Ma:2010dv}.
As shown in Fig.~\ref{fig:e2}(a), the event-averaged $\varepsilon_2$
from the normal AMPT model shows an overall decrease with $N_{\rm ch}$, while the  3-quark AMPT model shows a rather flat $\varepsilon_2$ versus $N_{\rm ch}$~\cite{Zheng:2021jrr}.
This helps to explain why $c_{2} \{2 \}$ from the normal AMPT model often shows a peak at moderate multiplicities in Figs.~\ref{fig:c22} and \ref{fig:c22d}. 

Since flows are directly related to the initial eccentricity, one can expect the flow fluctuations to depend on the  eccentricity fluctuations:
\begin{eqnarray}
c_{\varepsilon_{n}} \{2  \}= \langle \varepsilon_{n}^{2}  \rangle, 
~c_{\varepsilon_{n}} \{4  \}=\langle \varepsilon_{n}^{4}  \rangle-2 \langle  \varepsilon_{n}^{2}  \rangle ^{2},
\label{eq:ce24}
\end{eqnarray}
where the bracket represents the averaging over all events. 
If one assumes $v_{2}\propto \varepsilon_2$ event-by-event~\cite{Bhalerao:2006tp,Ma:2016hkg}, 
$c_{2} \{4  \}$ would be proportional to $c_{\varepsilon_{2}} \{4  \}$. 
Figure~\ref{fig:e2}(b) shows $c_{\varepsilon_2} \{4  \}$ 
versus the charged particle multiplicity, where the normal AMPT behaves very differently 
from the 3-quark AMPT with both models using  $\sigma=3$ mb.
We see some similarities between the $N_{\rm ch}$ dependences of 
$c_{\varepsilon_{2}} \{4  \}$ here and $c_{2} \{4  \}$ of Fig.~\ref{fig:c24}. 
For example, at $N_{\rm ch}(p_{\rm T}>0.4{\rm~GeV}/c) \in (20,70)$ 
both $c_{\varepsilon_{2}} \{4  \}$ and $c_{2} \{4  \}$ are negative from the normal AMPT 
while both are positive from the 3-quark AMPT. 
On the other hand, the development of $c_{2} \{4  \}$ depends not only 
on $c_{\varepsilon_{2}} \{4  \}$ but also sensitively on the parton interactions such as $\sigma$ 
as shown in Fig.~\ref{fig:sigma}.

\section{Conclusions}
\label{summary}

We have studied multiparticle cumulants including $c_{2} \{2 \}$ and $c_{2} \{4 \}$ in $p+p$ collisions at 13 TeV with a multi-phase transport model. 
Both the normal string melting version of the AMPT model and a 3-quark version that includes the proton sub-nucleon geometry 
are found to produce negative $c_{2} \{4 \}$ values 
at high multiplicities. 
We also find that both $c_{2} \{2 \}$ and $c_{2} \{4 \}$ 
depend strongly on the parton cross section $\sigma$ and the 
dependences at high multiplicities are non-monotonous. 
Furthermore, the $c_{2} \{4 \}$ value from the standard cumulant method 
is negative only when $\sigma$ is within a limited range, approximately [0.15, 1.5] mb. Quantitatively, however, none of the model results with a constant $\sigma$ can well reproduce the ATLAS $c_{2} \{4 \}$ data versus the charged particle multiplicity, and no model results can well reproduce the $c_{2} \{2 \}$ data with and without the pseudorapidity gap. Further studies are therefore needed for quantitative descriptions of these observables in high energy $p+p$ collisions.

Nevertheless, the results from the AMPT model share many qualitative features as the experimental data. 
They include the $c_{2} \{4 \}$ sign change from positive at low multiplicities 
to negative at high multiplicities for the standard cumulant method and default $p_{\rm T}$ range, which demonstrates the importance of nonflow effects at low multiplicities and also indicates a collective behavior at high multiplicities. 
They also include the decrease of $c_{2} \{4 \}$ when a higher $p_{\rm T}$ cut is applied to charged particles in the calculation of $N_{\rm ch}^{\rm sel}$ used for the event averaging, and the decrease of $c_{2} \{4 \}$ when the three-subevent cumulant method is used. 
In addition, we find that the results from the 3-quark AMPT model are in better agreements with the experimental data than the normal AMPT model, for example, in the shape of the multiplicity dependence of $c_{2} \{4 \}$. 
This indicates the importance of including the sub-nucleon geometry of the proton  in studies of multiparticle cumulants in $p+p$ collisions.

\begin{acknowledgments}
We thank Dr. Mao-Wu Nie for helping with the three-subevent method, Dr. You Zhou and Liu-Yao Zhang for helpful discussions, and Dr. Chen Zhong for maintaining the high-quality performance of Fudan supercomputing platform for nuclear physics. This work is supported in part by the National Natural Science Foundation of China under Grant Nos. 11961131011, 11890714, 11835002, 12105054 (X.-L.Z. and G.-L.M.) and 11905188 (L.Z.), the China Postdoctoral Science Foundation under Grant No. 2021M690708 (X.-L.Z.), the National Science Foundation under Grant No. PHY-2012947 (Z.-W.L.), the Strategic Priority Research Program of Chinese Academy of Sciences under Grant No. XDB34030000 (G.-L.M.), and the Guangdong Major Project of Basic and Applied Basic Research under Grant No. 2020B0301030008 (G.-L.M.). 
\end{acknowledgments}

\end{document}